\documentclass[aps,pra,onecolumn,notitlepage,superscriptaddress,nofootinbib]{revtex4-2}

\usepackage[T1]{fontenc}
\usepackage{lmodern}
\usepackage{graphicx}
\usepackage{amsmath,amssymb}
\usepackage{amsthm}
\usepackage{physics}
\usepackage[colorlinks=true,linkcolor=blue,citecolor=blue,urlcolor=blue]{hyperref}
\usepackage{xcolor}
\usepackage{caption}
\usepackage{bm} % For bold math
\usepackage{graphicx}
\usepackage{caption}
\usepackage{subcaption}  % برای کپشن جدا برای هر تصویر
\usepackage{amsmath}
\usepackage{placeins}
\newcommand{\pt}{$\mathcal{P}$\,$\mathcal{T}$}
\newcommand{\APT}{Anti\,-\,$\mathcal{P}$\,$\mathcal{T}$}

\newcommand{\be}{\begin{equation}}
	\newcommand{\ee}{\end{equation}}
\newcommand{\ba}{\begin{eqnarray}}
	\newcommand{\ea}{\end{eqnarray}}

\begin{document}

	\title{	Quantum Dynamics and Information Measures in \pt\ and \APT-Symmetric Systems}
	
	\author{Amir Ahmadi}
	\email{Amir.Ahmadi@email.kntu.ac.ir}
	\affiliation{Department of Physics, K.N. Toosi University of Technology, P.O. Box 15875-4416, Tehran, Iran}

	\author{Roozbeh H. Asgari}
	\email{roozbeh.hosseinasgari@email.kntu.ac.ir}
	\affiliation{Department of Physics, K.N. Toosi University of Technology, P.O. Box 15875-4416, Tehran, Iran}

	\author{Javad T. Firouzjaee} 
	\email{firouzjaee@kntu.ac.ir}
	\affiliation{Department of Physics, K.N. Toosi University of Technology, P.O. Box 15875-4416, Tehran, Iran}
	\affiliation{School of Physics, Institute for Research in Fundamental Sciences (IPM), P.O. Box 19395-5531, Tehran, Iran}
	\affiliation{PDAT Laboratory, Department of Physics, K.N. Toosi University of Technology, P.O. Box 15875-4416, Tehran, Iran}
	
	\date{\today}

	\date{\today}
	\begin{abstract}
		
		In this study, we investigate  qubit dynamics under \pt  and \APT-symmetric non-Hermitian Hamiltonians, focusing on phase evolution, decoherence, quantum speed limits (QSL), and Rényi entanglement entropies. Using similarity transformations and Dyson maps, we analyze the reduced density matrix evolution in bosonic environments. Anti-\(\mathcal{PT}\)-symmetric systems show enhanced robustness against decoherence, with slower entropy growth and longer coherence times compared to \(\mathcal{PT}\)-symmetric counterparts. QSL behavior is non-monotonic, reflecting rapid initial evolution followed by gradual decrease. Higher-order Rényi entropies reveal that \APT-symmetric qubits preserve quantum information more effectively, offering advantages for memory and cryptographic applications.
	\end{abstract}

	\maketitle
	\tableofcontents

	\section{Introduction}
	
	The exploration of quantum systems beyond the traditional Hermitian framework has attracted substantial interest in recent years, motivated by the recognition that non-Hermitian Hamiltonians can generate physically meaningful and often novel dynamics. A key development in this direction was the realization that non-Hermitian systems exhibiting \textit{parity-time} (\pt) symmetry can still possess entirely real spectra, as first shown by Bender and Boettcher~\cite{bender1998real}. This pivotal insight established that Hermiticity, while sufficient, is not necessary for real eigenvalues, thus launching the field of \pt-symmetric quantum mechanics.
	
	These symmetry conditions are particularly relevant in open quantum systems, where non-Hermiticity naturally arises due to environmental coupling~\cite{breuer2002theory}. \pt-symmetric systems display a well-known phase transition between an \textit{unbroken} phase, with real eigenvalues and quasi-unitary evolution, and a \textit{broken} phase, where eigenvalues appear in complex-conjugate pairs and probability is not conserved~\cite{mostafazadeh2005pt,mostafazadeh2007time}. In \APT-symmetric systems, the dynamics are inherently dissipative but exhibit distinct coherence patterns that can be harnessed in engineered quantum setups~\cite{cen2022,wang2022fisher,junjie}.
	
	Recent research has unveiled a broad spectrum of applications for \pt{} and \APT-symmetry in diverse quantum platforms such as quantum computing ~\cite{gardas2016pt,nielsen2010quantum,xu2024ptmapping,el2018non}, quantum sensing ~\cite{miri2019exceptional,wang2022fisher,zhang2023sensing,hodaei2017enhanced}, quantum optics and photonics ~\cite{shang,ozdemir2019non,choi2018observation,miao2024enhanced,guo2009observation}, quantum thermodynamic ~\cite{Deffner2013,Deffner2017,funo2023thermo}, quantum chaos and scrambling ~\cite{li2021information,bhattacharya2020scrambling} and topological quantum systems ~\cite{gong2018topological,yao2018edge}.

	The role of \pt{} and \APT{} symmetry is especially profound in the context of qubit systems interacting with bosonic environments. In these systems—often modeled as spin-boson~\cite{breuer2002theory} or Jaynes-Cummings-like extensions—the presence of symmetry can dramatically alter decoherence rates, entanglement behavior, and information scrambling~\cite{cen2022}. The interplay between symmetry, dissipation, and entropic measures opens new pathways for quantum information preservation and control.
	
	In this work, we investigate the phase dynamics, quantum speed limits, and the evolution of various orders of Rényi entropies in both \pt and \APT-symmetric qubit systems. By developing a unified formalism for these two symmetry regimes, we aim to reveal how underlying non-Hermitian structures govern observable behavior—and how these effects may be harnessed in the design of next-generation quantum technologies.

	\section{Preliminaries}
	
	In this section, we introduce essential frameworks that underpin the study of \pt (Parity-Time) \cite{bender1998real,gardas2016pt,mostafazadeh2005pt,ptbhat}
	and \APT (Anti-Parity-Time) symmetry \cite{cen2022}, the spin-boson model \cite{breuer2002theory}, and the dephase and decoherence functions \cite{cen2022}. These concepts together shape the formalism and dynamics of our system, the two-level non-Hermitian qubit weakly coupled to a bosonic bath \cite{breuer2002theory}.\\
	
	\subsection{The Parity Operator \( \mathcal{P} \) and the Time-Reversal Operator \( \mathcal{T} \) }
	
	The parity operator performs spatial inversion and acts on position and momentum operators as:
	\begin{align}
		\mathcal{P} x \mathcal{P}^{-1} &= -x, \label{eq:P_x}\\
		\mathcal{P} p \mathcal{P}^{-1} &= -p. \label{eq:P_p}
	\end{align}
	It is a linear, Hermitian, and involutive operator:
	\begin{equation}
		\mathcal{P} = \mathcal{P}^\dagger, \qquad \mathcal{P}^2 = I. \label{eq:P_herm_invol}
	\end{equation}
	The parity operator's role in quantum systems extends beyond just spatial inversion; it also plays a fundamental part in the symmetry properties of the system's Hamiltonian \cite{sakurai1995modern}. The parity operation is critical in determining the system's response to spatial transformations and in categorizing quantum states into even and odd parity states. The involutive property of the parity operator ensures that performing the parity operation twice yields the identity, reinforcing its fundamental symmetry nature.\\
	
	The time-reversal operator reverses the sign of momentum and complex conjugates the imaginary unit:
	\begin{align}
		\mathcal{T} x \mathcal{T}^{-1} &= x, \label{eq:T_x}\\
		\mathcal{T} p \mathcal{T}^{-1} &= -p, \label{eq:T_p}\\
		\mathcal{T} i \mathcal{T}^{-1} &= -i. \label{eq:T_i}
	\end{align}
	Unlike \( \mathcal{P} \), the operator \( \mathcal{T} \) is anti-linear and anti-unitary. It generally satisfies:
	\begin{equation}
		\mathcal{T}^2 = I. \label{eq:T_sq}
	\end{equation}
	The time-reversal operator plays an essential role in reversing the flow of time in quantum systems, providing a symmetry that is particularly useful in studying systems out of equilibrium~\cite{wigner1959group}. It leads to profound implications in the dynamics of quantum systems, especially in scenarios where time-reversal symmetry is spontaneously broken. This symmetry also plays a critical role in quantum information theory, where the invariance of the system under time-reversal is a key factor in ensuring certain information-theoretic properties~\cite{viola2022time}.
	
	\subsection{\pt\ Symmetric and \APT\ Symmetric Qubit}
	
	\textbf{\pt-Symmetric Hamiltonians:}
	A Hamiltonian is said to be \pt-symmetric if it commutes with the combined operator \( \mathcal{PT} \) \cite{gardas2016pt,cen2022}:
	\begin{equation}
		[\mathcal{PT}, H] = 0 \quad \Leftrightarrow \quad \mathcal{PT} H (\mathcal{PT})^{-1} = H. \label{eq:PT_commute}
	\end{equation}
	This condition does not imply Hermiticity:
	\begin{equation}
		H \neq H^\dagger, \label{eq:H_nonherm}
	\end{equation}
	yet, under certain conditions, \( H \) can still have a real energy spectrum.
	
	\textbf{Unbroken and Broken \( \mathcal{PT} \)-Symmetry:}
	If an eigenfunction \( \psi \) of \( H \) is also an eigenfunction of \( \mathcal{PT} \), i.e.,
	\begin{equation}
		\mathcal{PT} \psi = \lambda \psi, \qquad |\lambda| = 1, \label{eq:PT_eigen}
	\end{equation}
	then \( \psi \) lies in the unbroken \pt{} phase, and the corresponding eigenvalue \( E \) is guaranteed to be real:
	\begin{equation}
		H \psi = E \psi, \qquad E \in \mathbb{R}. \label{eq:real_eigen}
	\end{equation}
	In this unbroken phase, the system exhibits normal quantum mechanical behavior, with real eigenvalues corresponding to stable energy states. On the other hand, when \pt-symmetry is spontaneously broken, the eigenvalues of the Hamiltonian become complex conjugates of each other, leading to a dramatic shift in the system's behavior~\cite{el2018non,bender2007making}. This phase transition between broken and unbroken \pt-symmetry is a fascinating phenomenon observed in various physical systems, such as optical cavities and quantum dots~\cite{el2018non}.
	
	\textbf{Anti-\pt-Symmetric Systems:}
	Beyond $\mathcal{PT}$-symmetric systems, recent work has introduced the notion of \emph{Anti-}$\mathcal{PT}$-symmetry, where the Hamiltonian $H$ satisfies an anti-commutation relation with the combined $\mathcal{PT}$ operator \cite{cen2022}. Specifically, an \APT-symmetric Hamiltonian obeys:
	\begin{equation}
		\{\mathcal{PT}, H\} = \mathcal{PT} H + H \mathcal{PT} = 0. \label{eq:APT_anticommute}
	\end{equation}
	
	The emergence of \APT-symmetry offers a new paradigm for studying systems with non-Hermitian dynamics, as it leads to entirely different spectral characteristics from those seen in \pt-symmetric systems.\\
	
	\subsection{Pseudo-Hermitian Systems and the Metric Operator}
	
	The notion of \pt-symmetry was later connected to the broader class of \emph{pseudo-Hermitian} Hamiltonians introduced by Mostafazadeh~\cite{mostafazadeh2005pt}. A non-Hermitian Hamiltonian \( H \) is called \emph{pseudo-Hermitian} if there exists a linear, invertible, and Hermitian operator \( \eta \) such that:
	\begin{equation}
		H^\dagger = \eta H \eta^{-1}. \label{eq:pseudo_herm}
	\end{equation}
	This relation implies that \( H \) is Hermitian with respect to a modified inner product:
	\begin{equation}
		\langle \psi | \phi \rangle_\eta := \langle \psi | \eta | \phi \rangle. \label{eq:eta_innerprod}
	\end{equation}
	If such a positive-definite \( \eta \) exists, then the time evolution governed by \( H \) is unitary in the \( \eta \)-inner product space~\cite{mostafazadeh2007time}. This allows for a consistent quantum theory even when \( H \neq H^\dagger \) in the usual sense.
	
	For many \pt-symmetric Hamiltonians, it is possible to construct such an \( \eta \) explicitly, showing that they are pseudo-Hermitian. In fact, the reality of the spectrum and the unbroken \pt-symmetry condition correspond to the existence of a positive-definite \( \eta \)~\cite{mostafazadeh2005pt,mostafazadeh2007time}.
	
	\textbf{Conditions for Reality of the Spectrum:}
	Thus, we may summarize the key conditions for a non-Hermitian Hamiltonian \( H \) to possess a real spectrum:
	\begin{enumerate}
		\item \( H \) is \pt-symmetric: \( [\mathcal{PT}, H] = 0 \) (see Eq.~\eqref{eq:PT_commute}),
		\item The \pt-symmetry is \emph{unbroken}: each eigenfunction \( \psi \) satisfies \( \mathcal{PT} \psi = \psi \) (cf. Eq.~\eqref{eq:PT_eigen}).
	\end{enumerate}
	Then, according to~\cite{gardas2016pt}, if the spectrum of the Hamiltonian is real, there exists a Hermitian, positive-definite operator \( \eta \) such that \( H^\dagger = \eta H \eta^{-1} \) (Eq.~\eqref{eq:pseudo_herm}).
	
	Although \pt\ and \APT\ systems have different physical properties and definitions, their dynamics are governed by the same framework of open quantum systems, which we introduce in the following subsections.
	
	\subsection{The Spin-Boson Model}
	
	We consider the spin-boson model as our fundamental open quantum system \cite{breuer2002theory}, which will later be extended to include \pt and \APT-symmetric dynamics. In this model, the reduced system is ultra weakly coupled to a bosonic bath, thus allowing only quantum information flow and no heat exchange between the system and the environment. This constraint leads to the famous \textit{pure decoherence} or \textit{dephase} phenomena\cite{gardas2016pt,cen2022}.
	
	The total Hamiltonian is expressed as:
	\begin{equation}
		H = H_S + H_B + H_I, \label{eq:H_total}
	\end{equation}
	where $H_S$ is the Hamiltonian of the reduced system, and
	\begin{equation}
		H_B = \sum_k \omega_k b^{\dagger}_{k} b_{k}, \label{eq:H_bath}
	\end{equation}
	is the bosonic bath Hamiltonian, and the bosonic creation and annihilation operators satisfy the canonical commutation relation:
	\begin{equation}
		[b_k, b^{\dagger}_{k'}] = \delta_{kk'}. \label{eq:b_commutation}
	\end{equation}
	
	The interaction Hamiltonian is generally taken as:
	\begin{equation}
		H_I = H_S \otimes V_B, \label{eq:H_int}
	\end{equation}
	with the bath operator defined as:
	\begin{equation}
		V_B = \sum_k \left( g_k b_k^{\dagger} + g_k^* b_k \right). \label{eq:V_B}
	\end{equation}
	Here, \(g_k\) represents the system-bath coupling strength for the \(k\)-th mode of the environment.

	\section{Dynamics: Phase Evolution and Decoherence}
	
	In open quantum systems, phase evolution and decoherence are two fundamental yet distinct phenomena that arise from interactions with the environment~\cite{breuer2002theory,lindblad1976generators}. 
	
	\textit{Decoherence} refers to the loss of quantum coherence between components of a superposition, which destroys interference without necessarily changing energy levels or populations. For instance, random fluctuations in a magnetic field or collisions with gas particles can turn a coherent quantum state into a classical mixture~\cite{gardas2016pt}. It manifests in the suppression of off-diagonal terms (coherences), signaling a transition from quantum to classical behavior~\cite{Deffner2013}.
	Physically, decoherence arises because the system becomes entangled with its environment, leaking quantum information into inaccessible degrees of freedom. 
	
	\textit{Phase evolution} also arises from the non-Hermitian, open system dynamics \cite{breuer2002theory}. We will show that it causes some phase amplification in the \pt-Symmetric system, while projecting phase decay in the \APT-symmetric case.
	
	Understanding and controlling these processes is vital for quantum technologies, where environmental coupling sets the ultimate limits for coherence times, entanglement preservation, and computational fidelity. In the following subsection, we introduce the non-Hermitian Hamiltonians of our \pt and \APT-symmetric qubit systems, their corresponding density matrices, and how phase evolution and coherence loss emerge within these dynamics.
	
	\vspace{3mm}
	\noindent
	\subsection{Hamiltonians for \pt \hspace{1mm} and \APT-Symmetric Qubits :}
	
	The general non-Hermitian system Hamiltonian in matrix form can be defined as\cite{cen2022}:
	\begin{equation} \label{eq:H_APT}
		H_S = 
		\begin{pmatrix}
			\alpha + i \theta   &   \xi + i \delta\\
			-\xi + i \delta   &   -\alpha + i \theta
		\end{pmatrix} \quad \text{(\APT-symmetric)}
	\end{equation}
	\begin{equation} \label{eq:H_PT}
		H_S^{\mathcal{PT}} = 
		\begin{pmatrix}
			\alpha + i \theta   &   \xi + i \delta\\
			\xi - i \delta   &   \alpha - i \theta
		\end{pmatrix} \quad \text{(\pt-symmetric)}
	\end{equation}
	
	where $\alpha,\, \theta,\, \delta,\, \xi \in \mathbb{R}$, with $\theta$ representing the amplification rate in \pt-symmetric systems and serving as a key parameter characterizing non-Hermitian behavior. In contrast, for \APT-symmetric systems, the parameters $\delta$ and $\xi$ govern the system's dynamical behavior.
	
	If we take the parity operator ${\mathcal{P}}$ as the Pauli matrix
	$\sigma_{x}$ and the time reversal operator ${\mathcal{T}}$ as its complex conjugate, we have both $[\mathcal{PT}, H_S^{\mathcal{PT}}]$ = 0 \hspace{1mm}and\hspace{1mm}$\{\mathcal{PT}, H_S\}$=0 \hspace{1mm}just\hspace{1mm}satisfied.\\
	
	Defining the transformation matrix:
	\begin{equation} \label{eq:T_matrix}
		T = 
		\begin{pmatrix}
			\omega_0 - \alpha & -\xi - i\delta \\
			\omega_0 + \alpha  &  \xi + i\delta
		\end{pmatrix},
	\end{equation}
	
	and using it for similarity transformation, we define:
	\begin{equation} \label{eq:omega_PT}
		\omega_0^{\mathcal{PT}} = \sqrt{\delta^2 + \xi^2 - \theta^2},
	\end{equation}
	which leads to a diagonalized Hermitian Hamiltonian $h_S^{D}$ such that:
	\begin{equation} \label{eq:h_diag_PT}
		h_S^{D} = T H_S^{\mathcal{PT}} T^{-1},
	\end{equation}
	with resulting eigenvalues given by:
	\begin{equation} \label{eq:eigvals_PT}
		E_{\pm} = \alpha \pm \omega_0^{\mathcal{PT}}.
	\end{equation}
	
	In this work, we focus on the regime where the eigenvalue splitting remains real. This corresponds to the parametric condition:
	\begin{equation} \label{eq:param_cond_PT}
		\delta^2 + \xi^2 \geq \theta^2.
	\end{equation}

	To diagonalize the total system Hamiltonian \( H \) for the \APT-symmetric case, we again employ the transformation from Eq.~\eqref{eq:T_matrix}, with the resulting eigenvalue:
	\begin{equation} \label{eq:omega_APT}
		\omega_0^{\mathcal{APT}} = \sqrt{\alpha^2 - \xi^2 - \delta^2}.
	\end{equation}
	
	The transformed diagonal Hamiltonian is:
	\begin{equation} \label{eq:H_diag_APT}
		H_D = T H T^{-1} = (-\omega_0^{\mathcal{APT}} \sigma_z + i\theta)(1 + V_B) + H_B.
	\end{equation}
	
	This diagonal form corresponds to the eigenvalue expansion:
	\begin{equation} \label{eq:eig_expansion}
		H = \sum_n E_n \ket{n} \bra{n},
	\end{equation}
	with \( E_n \in \mathbb{C} \). Thus, \( H \) can be recast using a biorthonormal eigenbasis:
	\begin{equation} \label{eq:eig_biorthogonal}
		H = \sum_n E_n \ket{\psi^R_n} \bra{\psi^L_n},
	\end{equation}
	where the right and left eigenvectors are defined as:
	\begin{equation} \label{eq:right_left_eig}
		\ket{\psi^R_n} = T^{-1} \ket{n}, \qquad \bra{\psi^L_n} = \bra{n} T,
	\end{equation}
	and satisfy the normalization and completeness relations:
	\begin{equation} \label{eq:normalization_completeness}
		\bra{\psi^L_n}\ket{\psi^R_m} = \delta_{nm}, \qquad \sum_n \ket{\psi^R_n} \bra{\psi^L_n} = I.
	\end{equation}
	
	The two eigenvalues of the \APT-symmetric Hamiltonian (Eq.~\eqref{eq:H_APT}) are given by:
	\begin{equation} \label{eq:eigvals_APT}
		E_{\pm} = i\theta \pm \omega_0^{\mathcal{APT}}.
	\end{equation}

	In the general case of a non-Hermitian Hamiltonian \( H \), prior studies~\cite{mostafazadeh2002pseudo,bender2007making,el2018non} have proposed expressing it as a sum of Hermitian and anti-Hermitian components:
	\begin{equation} \label{eq:H_herm_decomp}
		H = H_R + i H_I,
	\end{equation}
	where 
	\begin{equation} \label{eq:herm_parts}
		H_R = \frac{H + H^\dagger}{2} ,  \quad H_I = \frac{H - H^\dagger}{2i} 
	\end{equation}
	are Hermitian operators representing the real and imaginary parts of \( H \), respectively.

	\subsection{Dynamics of \pt-symmetric qubit}
	
	Since the \pt-symmetric Hamiltonian~\eqref{eq:H_PT} led to a diagonalized Hermitian one~\eqref{eq:h_diag_PT} under the similarity transformation~\eqref{eq:T_matrix}, we can obtain its dynamics using the conventional Hermitian-valid formalism.
	
	The reduced density matrix for the system is recovered as a partial trace over the environment of the overall system plus bath evolution. Subsequently, it can be expressed in terms of the reduced density matrix for the diagonalized system under normalization as
	\begin{equation} \label{eq:rhoS_t}
		\rho_S(t) = \frac{T^{-1} \rho^D_S(t) (T^{-1})^\dagger}{\text{Tr}_S \left[ T^{-1} \rho^D_S(0) (T^{-1})^\dagger \right]}
	\end{equation}
	
	\begin{equation} \label{eq:rhoD_0}
		\rho^D_S(0) = \begin{pmatrix}
			\rho^D_{11} & \rho^D_{12} \\
			\rho^D_{21} & \rho^D_{22}
		\end{pmatrix}
	\end{equation}
	
	\begin{equation} \label{eq:rhoD_t}
		\rho^D_S(t) = 
		\begin{pmatrix}
			\rho^D_{11} & 0 \\
			0 & \rho^D_{22}
		\end{pmatrix}
		+
		\begin{pmatrix}
			0 & \rho^D_{12}(t) \\
			\rho^D_{21}(t) & 0
		\end{pmatrix}
		e^{-(\omega_0^{\mathcal{APT}})^2 \gamma(t)}
	\end{equation}
	
	\begin{equation} \label{eq:rhoD_offdiag}
		\rho^D_{12}(t) = e^{2i \omega_0 t} e^{-i \omega_0^{\mathcal{APT}} \Omega(t)}, \quad 
		\rho^D_{21}(t) = \left[ \rho^D_{12}(t) \right]^*
	\end{equation}
	
	where
	\begin{equation} \label{eq:Omega_t}
		\Omega(t) = 4\theta \int_0^\infty dw \, J(w) \frac{w t - \sin(w t)}{w^2}
	\end{equation}
	
	and
	\begin{equation} \label{eq:gamma_t}
		\gamma(t) = 4 \int_0^\infty dw \, J(w) \frac{1 - \cos(w t)}{w^2} \coth \left( \frac{\beta w}{2} \right)
	\end{equation}
	
	and
	\begin{equation} \label{eq:spectral_density}
		J(w) = \sum_k |g_k|^2 \delta(w - w_k)
	\end{equation}
	
	are respectively the function influencing phase evolution, the decoherence factor, and the spectral density of the bath \cite{breuer2002theory,cen2022}.
	
	Realizing that the similarity transformation does not turn the \APT-symmetric Hamiltonian into a Hermitian one, so we cannot use the same dynamics formalism as that of the \pt system, we utilize a different, more powerful approach to calculate the dynamics of that system.

	\textbf{Time-Dependent Dyson Map:} 
	It has been shown that non-Hermitian Hamiltonians can be reformulated and re-investigated within the framework of Hermitian Hamiltonians using a time-dependent Dyson map as the key tool, as analyzed for example in \cite{cen2022} for general non-Hermitian systems, and previously in studies of\hspace{1mm}\pt-symmetric bosonic systems coupled to baths \cite{gardas2016pt} and \pt-symmetric Jaynes-Cummings Hamiltonians \cite{cen2022}. 
	
	We will utilize the approach developed in \cite{cen2022} for general non-Hermitian systems and then specify it for our purposes in\hspace{1mm}\pt\hspace{1mm}and \APT-symmetric systems.
	
	First, consider a Hermitian system characterized by a diagonal Hermitian Hamiltonian $h_D$, with an associated density matrix $\rho_{Dh}$. Putting $\hbar$=1, the standard Liouville–Von Neumann equation for this setup is:
	\be \label{eq:DysonHermitianLiouville}
	i\partial_t \rho_{Dh} = [h_D, \rho_{Dh}], 
	\ee
	which governs the unitary evolution of the Hermitian system \cite{nielsen2010quantum}.
	
	Turning to the non-Hermitian case, specifically the diagonalized non-Hermitian Hamiltonian $H_D$, one obtains a modified Liouville–Von Neumann equation by combining the Schrödinger equation and its Hermitian conjugate:
	\be \label{eq:DysonNonHermitianLiouville}
	i\partial_t \rho_D = H_D \rho_D - \rho_D H_D^\dagger. 
	\ee
	This equation governs the evolution of the density matrix in the non-Hermitian framework \cite{mostafazadeh2007time}.
	
	Assuming that $H_D$ and $h_D$ are connected through a time-dependent Dyson map $\eta$ via the relation:
	\be \label{eq:DysonRelation}
	H_D = \eta^{-1} h_D \eta - i\eta^{-1}(\partial_t \eta), 
	\ee
	where the map $\eta$ connects the eigenstates of the Hermitian and non-Hermitian systems as $\ket{\phi_i} = \eta \ket{\psi_i}$ \cite{mostafazadeh2002pseudo, cen2022}.
	
	Inserting this Dyson transformation into Eq.~\eqref{eq:DysonNonHermitianLiouville} and comparing with Eq.~\eqref{eq:DysonHermitianLiouville} yields a mapping between the density matrices:
	\be \label{eq:DensityMatrixRelation}
	\rho_{Dh} = \eta \rho_D \eta^\dagger.
	\ee
	With the spectral decomposition $\rho_{Dh} = \sum_i P_i \ket{\phi_i} \bra{\phi_i}$, it follows that
	\be \label{eq:NonHermitianSpectral}
	\rho_D = \sum_i P_i \ket{\psi_i} \bra{\psi_i},
	\ee
	indicating that the Dyson map preserves the probability distribution $\{P_i\}$ \cite{mostafazadeh2007time}.
	
	Conversely, starting from the Hermitian Liouville–Von Neumann equation \eqref{eq:DysonHermitianLiouville} and applying the Dyson relation \eqref{eq:DysonRelation} again, one obtains an alternative evolution equation:
	\be \label{eq:CommutatorForm}
	i\partial_t \tilde{\rho}_D = [H_D, \tilde{\rho}_D], 
	\ee
	where $\tilde{\rho}_D = \eta^{-1} \rho_{Dh} \eta = \rho_D M$, with the metric operator defined as $M = \eta^\dagger \eta$. This construction leads to a quasi-Hermitian Hamiltonian:
	\be \label{eq:QuasiHermitianH}
	H_Q = H_D + i \eta^{-1}(\partial_t \eta),
	\ee
	which satisfies the condition
	\be \label{eq:QuasiHermitianCondition}
	\bra{H_Q \psi_i} M \ket{\psi_i} = \bra{\psi_i} M H_Q \ket{\psi_i} = \bra{\phi_i} h_D \ket{\phi_i},
	\ee
	and also the quasi-Hermiticity relation $(H_Q)^\dagger M = M H_Q$ \cite{mostafazadeh2007time, cen2022}.

	\vspace{3mm}
	\noindent
	\textbf{Density Matrix Evolution of the \APT-symmetric Qubit:}

	We now determine the equivalent Hermitian system using a time-dependent Dyson map. Starting from the diagonalized non-Hermitian Hamiltonian from Eq.~\eqref{eq:H_diag_APT}, and choosing an ansatz for the Dyson map \cite{cen2022}:
	\begin{equation}
		\eta = \exp[\varphi(t)(1 + V_B)]
		\label{eq:APT_eta_ansatz}
	\end{equation}
	the transformed Hermitian Hamiltonian becomes:
	\begin{equation}
		h_D = (-\omega_0^{\mathrm{APT}} \sigma_z + i\theta)(1 + V_B) + H_B - \varphi \tilde{V}_B - \frac{\varphi^2}{2} \Omega_k + i \partial_t \varphi,
		\label{eq:APT_hd_general}
	\end{equation}
	where
	\[
	\tilde{V}_B = \sum_k \omega_k (g_k a_k^\dagger - g_k^* a_k), \quad \Omega_k = \sum_k \omega_k |g_k|^2.
	\]
	
	To ensure that $h_D$ remains Hermitian, we impose the constraint:
	\[
	\partial_t \varphi = -\theta,
	\]
	which allows us to take:
	\begin{equation}
		\eta = \exp[-\theta t (1 + V_B)]
		\label{eq:APT_eta_final}
	\end{equation}
	The resulting Hermitian Hamiltonian is then:
	\begin{equation}
		h_D = -\omega_0^{\mathrm{APT}} \sigma_z + H_B - \omega_0^{\mathrm{APT}} \sigma_z V_B + \theta t \tilde{V}_B - \frac{\theta^2 t^2}{2} \Omega_k.
		\label{eq:APT_hd_final}
	\end{equation}
	
	We can alternatively represent $h_D$ through a series expansion in terms of the quasi-Hermitian Hamiltonian $H_Q$:
	\begin{equation}
		h_D = \sum_{n=0}^\infty \frac{(-1)^n}{n!} C_G^{(n)}(H_Q),
		\label{eq:APT_comm_expansion_HQ}
	\end{equation}
	where $G = \theta t (1 + V_B)$ and
	\begin{equation}
		C_G^{(n)}(O) = [G, [G, \ldots [G, O] \ldots]]
		\label{eq:APT_nested_comm}
	\end{equation}
	denotes the $n$-fold nested commutator between $G$ and the operator $O$.
	
	Similarly, in terms of the original non-Hermitian Hamiltonian $H_D$, we can express $h_D$ as:
	\begin{equation}
		h_D = -i\theta (1 + V_B) + \sum_{n=0}^\infty \frac{(-1)^n}{n!} C_G^{(n)}(H_D)
		\label{eq:APT_comm_expansion_HD}
	\end{equation}
	
	To ensure the eigenvalue gap remains real, we constrain the system parameters to lie within the regime:
	\[
	\alpha^2 \geq \delta^2 + \xi^2.
	\]
	
	The final step is to compute the density matrix associated with Eq.~\eqref{eq:APT_hd_final}, from which we can extract the decoherence characteristics of our qubit system.
	
	We consider the qubit dynamics where the system is initially uncorrelated with the bath, which is in a thermal Gibbs state, i.e.,\cite{breuer2002theory}
	
	\[
	\Omega_B = \frac{e^{-\beta H_B}}{Z}, \quad Z = \Tr_B\left(e^{-\beta H_B}\right),
	\]
	
	so that the full system-bath density matrix at \( t = 0 \) is given by:
	\begin{equation}
		\rho^{Dh}(0) = \rho_S^{Dh}(0) \otimes \Omega_B
		\label{eq:APT_rho0_total}
	\end{equation}
	
	We assume a general form for the initial reduced density matrix of the qubit\cite{cen2022}:
	\begin{equation}
		\rho_S^{Dh}(0) = 
		\begin{pmatrix}
			\frac{1}{2}(1 + \langle \sigma_z \rangle) & \langle \sigma_- \rangle \\
			\langle \sigma_+ \rangle & \frac{1}{2}(1 - \langle \sigma_z \rangle)
		\end{pmatrix}
		\label{eq:APT_rho0_qubit}
	\end{equation}
	
	where \( \sigma_\pm = \sigma_x \pm i\sigma_y \). The time-evolved reduced density matrix \( \rho_S^{Dh}(t) \) can then be obtained by evaluating the time-dependent expectation values of the Pauli operators, i.e., \( \sigma_z(t), \sigma_\pm(t) \).
	
	The expectation value of a general operator \( O(t) \) in this setup with a Hermitian Hamiltonian \( h_D \) is given by:
	\begin{equation}
		\langle O(t) \rangle = \Tr\left[ O(t) \rho(0) \right], \quad \text{where } O(t) = e^{i \int h_D dt} O(0) e^{-i \int h_D dt}
		\label{eq:APT_Ot_expect}
	\end{equation}
	
	To compute the decoherence behavior, we begin with the bath operators. Using the commutation relations \( [H_B, a_k] = -\omega_k a_k \), \( [V_B, a_k] = -g_k \), and \( [\tilde{V}_B, a_k] = -\omega_k g_k \), we obtain the equation of motion:
	\begin{equation}
		\partial_t a_k = -i\omega_k \left[ a_k - \omega_0^{\mathcal{APT}} \frac{g_k}{\omega_k} \sigma_z + \theta g_k t \right]
		\label{eq:APT_ak_eq}
	\end{equation}
	
	which leads to the solution for the time-dependent bath operator:
	\begin{equation}
		a_k(t) = -\theta g_k t + e^{-i\omega_k t} \left[ a_k - A_k(t) \sigma_z + B_k(t) \right]
		\label{eq:APT_akt}
	\end{equation}
	
	with
	\begin{equation}
		A_k(t) = \omega_0^{\mathcal{APT}} \frac{g_k}{\omega_k} (1 - e^{i\omega_k t})
		\label{eq:APT_Ak}
	\end{equation}
	
	\begin{equation}
		B_k(t) = i \theta \frac{g_k}{\omega_k} (1 - e^{i\omega_k t})
		\label{eq:APT_Bk}
	\end{equation}
	
	Using similar methods, the time evolution of the qubit operators is given by:
	\begin{equation}
		\sigma_z(t) = \sigma_z
		\label{eq:APT_sigmaz_t}
	\end{equation}
	
	\begin{equation}
		\sigma_\pm(t) = e^{\mp 2i \omega_0^{\mathcal{APT}} t} \, \mathcal{T}_+ \exp\left[ \int_0^t d\tau \, (\mp 2i \omega_0^{\mathcal{APT}}) \sum_k \left( g_k a_k^\dagger(\tau) + g_k^* a_k(\tau) \right) \right] \sigma_\pm
		\label{eq:APT_sigmapm_t}
	\end{equation}
	
	The expectation values of the qubit operators become:
	\begin{equation}
		\langle \sigma_z(t) \rangle = \langle \sigma_z \rangle
		\label{eq:APT_sz_expect}
	\end{equation}
	
	\begin{equation}
		\langle \sigma_\pm(t) \rangle = \langle \sigma_\pm \rangle \, e^{\mp 2i \omega_0^{\mathcal{APT}} t} \, e^{\pm i \omega_0^{\mathcal{APT}} [\Omega_2(t) - \Omega_1(t)]} \, e^{-(\omega_0^{\mathcal{APT}})^2 \gamma(t)}
		\label{eq:APT_spm_expect}
	\end{equation}
	
	Using $\ket{\psi}=\dfrac{(\ket{0}+\ket{1})}{\sqrt{2}}$ as the basis for calculating the expectation values at t=0, the reduced density matrix for the system under \APT-symmetry evolves as:
	\begin{equation}
		\rho_s^{Dh}(t) =
		\begin{pmatrix}
			\rho_{11}^{Dh} & 0 \\
			0 & \rho_{22}^{Dh}
		\end{pmatrix}
		+
		\begin{pmatrix}
			0 & \rho_{12}^{Dh}(t) \\
			\rho_{21}^{Dh}(t) & 0
		\end{pmatrix}
		e^{-(\omega_0^{\mathcal{APT}})^2 \gamma(t)}
		\label{eq:APT_rho_evolved}
	\end{equation}
	
	The coherence terms evolve as:
	\begin{equation}
		\rho_{12}^{Dh}(t) = e^{2i \omega_0^{\mathcal{APT}} t} \cdot e^{-i \omega_0^{\mathcal{APT}} \left[\Omega_2(t) - \Omega_1(t)\right]}, \quad
		\rho_{21}^{Dh}(t) = \left[\rho_{12}^{Dh}(t)\right]^*
		\label{eq:APT_coherence_terms}
	\end{equation}

	Adopting the spectral density function from Eq.~\eqref{eq:gamma_t} and using the Feynman-Vernon influence functional formalism , the auxiliary phase evolution functions for the system-environment interaction are given by:
	\begin{equation}
		\Omega_1(t) = 4\theta \int_0^\infty \frac{d\omega\, J(\omega)\left(1 - \cos(\omega t)\right)}{\omega^2}
		\label{eq:APT_Omega1}
	\end{equation}
	\begin{equation}
		\Omega_2(t) = 2\theta t^2 \int_0^\infty d\omega\, J(\omega)
		\label{eq:APT_Omega2}
	\end{equation}
	
	Also, again taking Eq.~\eqref{eq:gamma_t} as the dephase factor, the decoherence function reads
	\begin{equation}
		D(t) = e^{-(\omega_0^{\mathcal{APT}})^2 \gamma(t)}
		\label{eq:APT_decoherence}
	\end{equation}
	
	\vspace{3mm}
	\noindent
	\subsection{Decoherence and phase evolution for both systems}
\begin{figure}[httb]
		\centering
		% عکس اول
		\begin{subfigure}[b]{0.30\textwidth}
			\includegraphics[width=\textwidth]{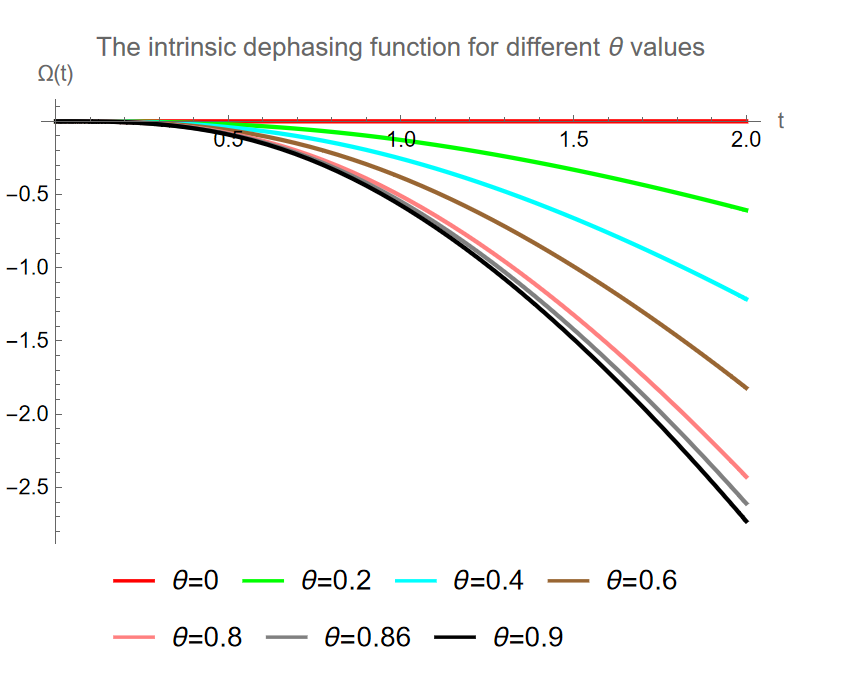}
			\caption{Phase evolution function}
			\label{fig:image10}
		\end{subfigure}
		\hfill
		% عکس دوم
		\begin{subfigure}[b]{0.35\textwidth}
			\includegraphics[width=\textwidth]{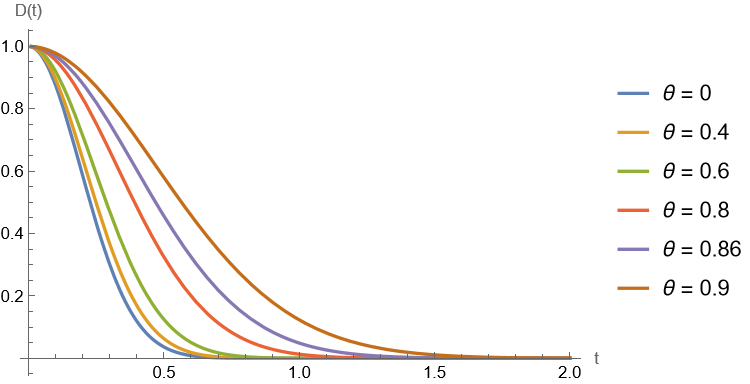}
			\caption{Decoherence function}
			\label{fig:pt_image2}
		\end{subfigure}
		
		\caption{Parameter values for the \pt-Symmetric Qubit: \( J_0 = 1,\ \beta = 0.5,\ \omega_c = 1,\ \mu = -0.5,\ \xi = 0.81,\ \alpha = 1,\ \delta = 0.56 \). }
		\label{fig:gamma_two}
	\end{figure}

	For the \pt-symmetric qubit,Fig.~\ref{fig:image10}  indicates that the phase evolution function is negative, monotonically decreasing and $\theta$-dependent, thus implying amplification except for $\theta=0$ or Hermitian case. As the value of $\theta$ increases, system tends to amplify more drastically. On the other hand, as shown in Fig.~\ref{fig:pt_image2}, the decoherence function is also $\theta$-dependent, and it exhibits smoother behavior for larger values of $\theta$; therefore, confirming slower decoherence as the non-Hermitian term increases.
	
\begin{figure}[httb]
		\centering
		% عکس اول
		\begin{subfigure}[b]{0.30\textwidth}
			\includegraphics[width=\textwidth]{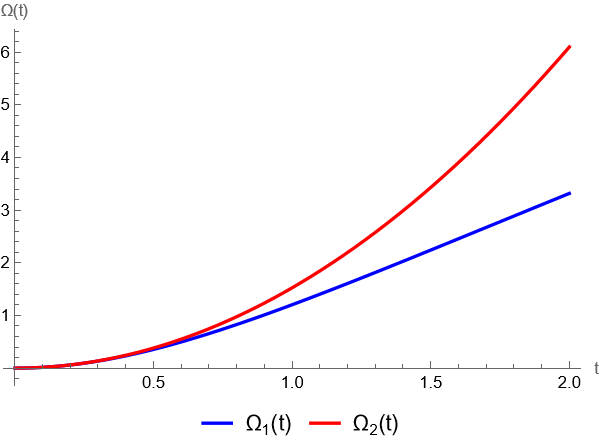}
			\caption{Phase evolution function}
			\label{fig:aptdephase}
		\end{subfigure}
		\hfill
		% عکس دوم
		\begin{subfigure}[b]{0.35\textwidth}
			\includegraphics[width=\textwidth]{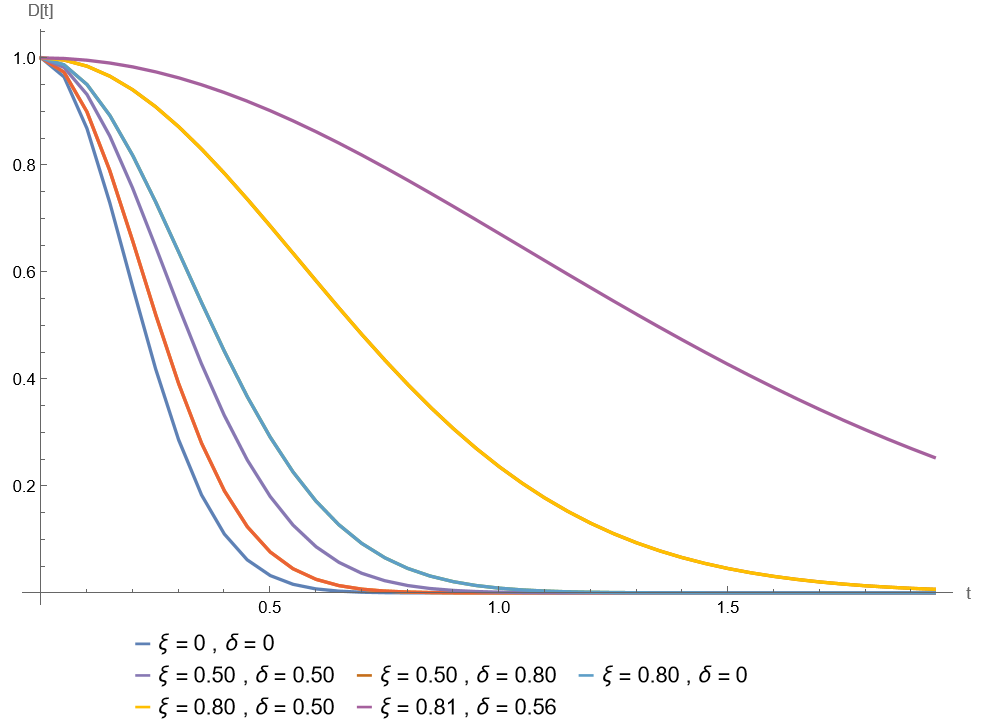}
			\caption{Decoherence}
			\label{fig:apt_image2}
		\end{subfigure}
		
		\caption{Parameter values for the \APT-symmetric Qubit: $J_0 = 1$, $\beta = 0.5$, $\Omega_c = 1$, $\mu = -0.5$, $\theta = 0.86$, $\alpha = 1$. }
		\label{fig:aptdecoh}
	\end{figure}

	\APT-symmetric system behaves more smoothly and generally lose less information compared to \pt-symmetric systems. For instance; at $\xi$ = 0.81, $\delta$ =  0.56 and $\theta$ = 0.86~\cite{cen2022}, \pt-symmetric system has dephased a lot faster than anti-\pt-symmetric system, Fig.~[\ref{fig:pt_image2} , \ref{fig:apt_image2}]. Also, in anti-\pt-symmetric systems, increase of non-Hermitian terms leads to slower decoherence. Fig.~\ref{fig:aptdephase} indicates that the phase evolution of the \APT-symmetric systems is independent of $\xi$ and $\delta$ and does not change as the values of the terms differ.

	%J0 = 1;
	%\[Beta] = 0.5;
	%\[Omega]c = 1;
	%\[Mu] = -0.5;
	%\[Xi] = 0.81;
	%\[Alpha] = 1;
	%\[Delta] = 0.56;

	\FloatBarrier

	\vspace{3mm}
	\section{Quantum Speed Limit for the\hspace{2mm}\pt\hspace{1mm}and \APT-symmetric\hspace{1mm}qubits}

	\subsection{Quantum Speed Limit}
	
	\textbf{Quantum Speed Limit, The Mandelstam–Tamm Bound:}
	In our investigation of the non-Hermitian dynamics, the quantum speed limit (QSL) is evaluated using the Mandelstam–Tamm-type bound. The Mandelstam–Tamm (MT) bound sets a fundamental lower limit on the time it takes for a quantum system to evolve between two distinguishable states. Originally derived for closed Hermitian systems, it connects the minimal evolution time to the inverse of the energy uncertainty in the system, thus reflecting a deep energy-time uncertainty relation.
	
	As shown by Mandelstam and Tamm, and further generalized in various contexts \cite{mandelstam1945uncertainty, giovannetti2003quantum}, the MT bound can be derived by bounding the time-dependent fidelity (or overlap) between the initial and evolved quantum states. Using trigonometric inequalities and spectral decomposition of the Hamiltonian, one obtains a bound on the square modulus of the survival amplitude, which leads to a minimal evolution time inversely proportional to the standard deviation of energy.
	
	This framework has also been extended to mixed states and time-dependent (driven) dynamics using the Bures angle as a distance measure \cite{braunstein1994statistical, toth2016quantum}. In these generalizations, the QSL is expressed in terms of the time-averaged energy variance, offering a rigorous limit even in open systems.
	
	\textbf{Non-Markovian Dynamics and Quantum Speed Limit:}
	
	Deffner and Lutz (2013) analyzed the quantum speed limit for non-Markovian systems, where the dynamics are influenced by memory effects of the environment. In non-Markovian dynamics, the rate of change of the system's state is modified by the history of previous interactions, and as a result, the quantum speed limit can be slower than in Markovian systems. This work generalized the concept of quantum speed limits, incorporating these memory effects by modifying the expression for \( \tau_{QSL} \) and introducing a non-Markovian correction term that depends on the past evolution of the system's state. 
	
	In their framework\cite{Deffner2013}, the time evolution of the system state is characterized by a generalized Liouville superoperator, and the fidelity \( F(\rho(0), \rho(t)) \) is a central quantity in computing the quantum speed limit. This fidelity captures both the instantaneous evolution and the influence of the system's past states, thus reflecting the memory inherent in non-Markovian dynamics. 
	
	In our non-Markovian system, we adopt the Mandelstam–Tamm bound as derived from the time-local evolution of the reduced system's coherence terms. This approach enables us to track how the intrinsic decoherence mechanisms—arising from $\mathcal{PT}$ or \APT-symmetry affect the minimal time scale of state evolution.
	The quantum speed limit (QSL) provides an upper bound on the speed of a quantum system, evolving from one state to another. In the general framework suggested in \cite{Deffner2017}, the quantum speed limit is given by the expression:
	
	\begin{equation}
		V_{QSL} = \frac{\left\|L(\rho(t))\right\|_{\text{op}}}{2 \sin \mathcal{L} \cos \mathcal{L}}
	\end{equation}
	\begin{equation}
		\tau_{QSL} = \dfrac{(\sin \mathcal{L})^2}{\dfrac{1}{\tau} \int_0^\tau dt\, \left\|L(\rho(t))\right\|_{\text{op}}}
	\end{equation}
	
	where \( L(\rho(t)) \) is the Liouville superoperator that governs the time evolution of the quantum state, and \( \|\cdot\|_{\text{op}} \) denotes the operator norm.

	The angle \( \mathcal{L} \) is defined as the arccosine of the fidelity between the initial state \( \rho(0) \) and the state at time \( t \), \( \rho(t) \), which is given by:
	
	\begin{equation}
		\mathcal{L}(\rho(0), \rho(t)) = \arccos \sqrt{F(\rho(0), \rho(t))}
	\end{equation}
	\begin{equation}
		F(\rho(0), \rho(t)) = \left[\text{Tr}\left(\sqrt{\sqrt{\rho(0)}\rho(t)\sqrt{\rho(0)}}\right)\right]^2
	\end{equation}
	
	Here, \( F(\rho(0), \rho(t)) \) is the fidelity function, which measures the overlap between the initial and final quantum states.

	\subsection{QSL in \texorpdfstring{$\mathcal{PT}$}{PT}-Symmetric Systems}
	
	In this subsection, we analyze the behavior of the quantum speed limit velocity $V_{\text{QSL}}(t)$ in $\mathcal{PT}$-symmetric systems under varying values of the non-Hermitian parameter $\theta$. Fig.~\ref{fig:image1} displays the QSL for $\theta \in \{0, 0.2, 0.4, 0.6, 0.8, 0.86, 0.9\}$.
	
	At early times, all curves exhibit relatively high QSL values, indicating that the system can evolve rapidly from its initial state. As time progresses, $V_{\text{QSL}}(t)$ decreases monotonically, eventually approaching a plateau or minimal value. This decay behavior is a hallmark of decoherence or damping commonly associated with $\mathcal{PT}$-symmetric open quantum systems.
	
	A key feature of this regime is that for larger $\theta$ values (such as $0.86$ and $0.9$), the rise and fall of QSL is slower and much more smoother. This suggests that increasing $\theta$ enhances the system’s capacity for maintaining dynamical evolution before reaching a quasi-stationary state. Such behavior reflects the role of the $\mathcal{PT}$ symmetry in balancing gain and loss, thereby preserving coherence over extended periods.

	\subsection{QSL in Anti-{$\mathcal{PT}$}-Symmetric Systems}
\begin{figure}[htbp]
		\centering
		% عکس اول
		\begin{subfigure}[b]{0.40\textwidth}
			\includegraphics[width=\textwidth]{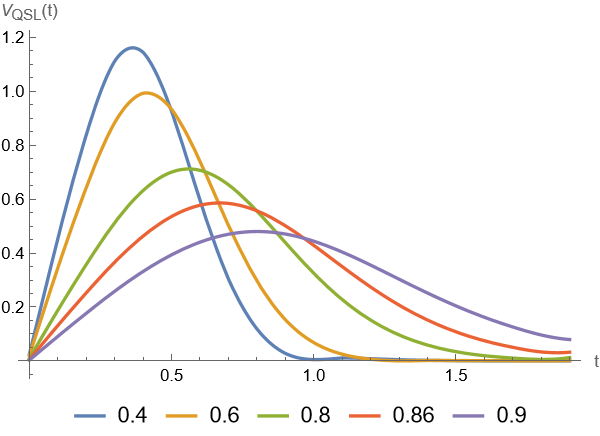}
			\caption{QSL in $\mathcal{PT}$-Symmetric System}
			\label{fig:image1}
		\end{subfigure}
		\hfill
		% عکس دوم
		\begin{subfigure}[b]{0.55\textwidth}
			\includegraphics[width=\textwidth]{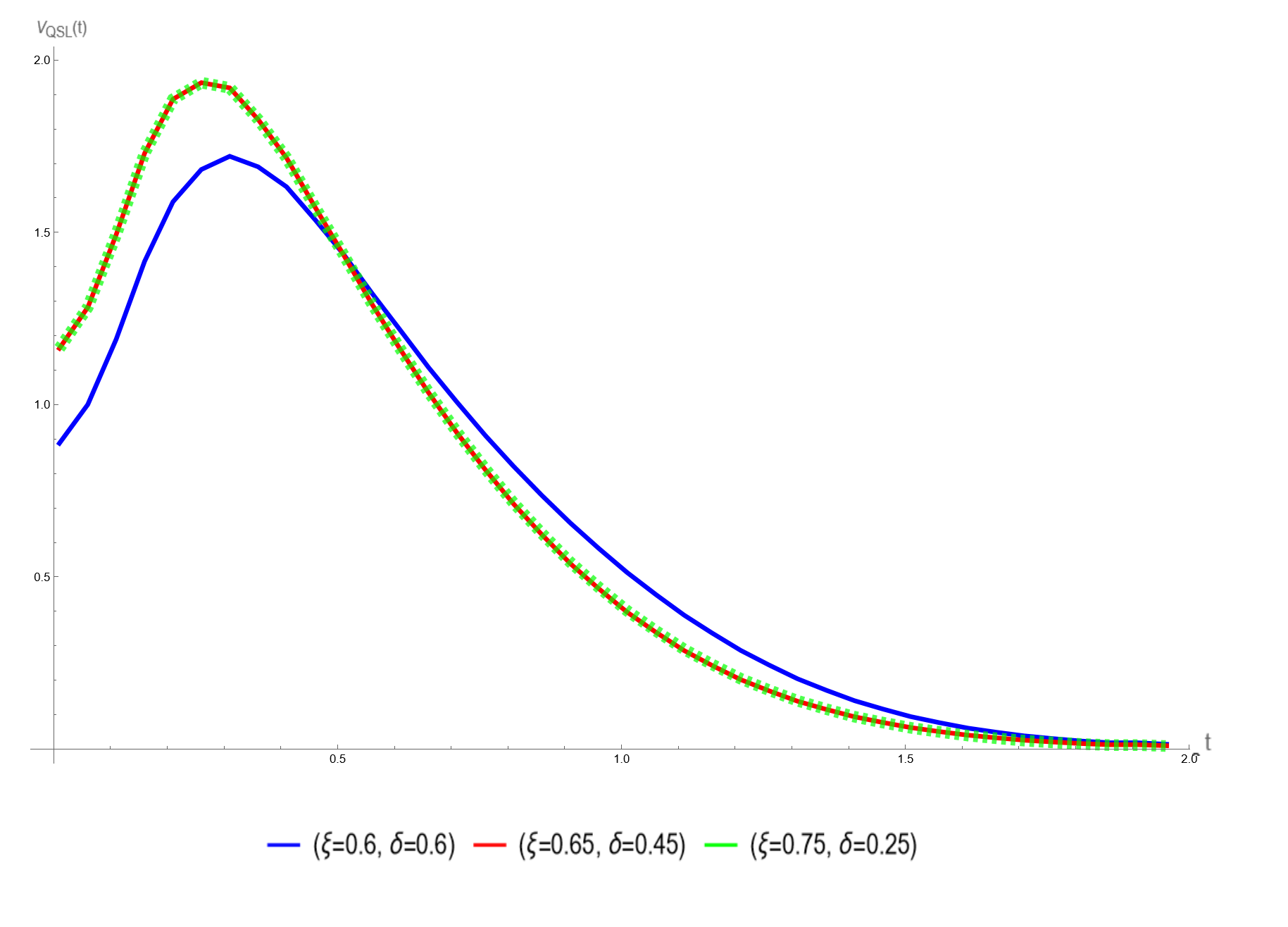}
			\caption{QSL in Anti-$\mathcal{PT}$-Symmetric System}
			\label{fig:image2}
		\end{subfigure}
		
		\caption{System parameters used for the simulations, for \pt-symmetric system: $J_0 = 1$, $\beta = 0.5$, $\Omega_c = 1$, $\mu = -0.5$, $\xi = 0.81$, $\alpha = 1$, $\delta = 0.56$, and for \APT-symmetric system: $J_0 = 1$, $\beta = 0.5$, $\Omega_c = 1$, $\mu = -0.5$, $\theta = 0.86$, $\alpha = 1$.}
		
		\label{fig:sidebyside}
	\end{figure}

	As shown in Fig.\ref{fig:image2} , for the
	\APT-symmetric system, we see similar trends of initial increase and final slowing down of the quantum speed limit. Surprisingly, some pairs of the non-Hermitian terms tend to have similar behavior and curves, as for the pairs ($\xi$ = 0.65, $\delta$ = 0.45) and ($\xi$ = 0.75, $\delta$ = 0.25). 
	Also, we can see that increasing only one of the non-Hermitian terms ($\xi$) while decreasing the other ($\delta$) doesn't allow the system to slow down, while instead keeping the parameters close to each other causes better and slower decoherence, although the rise and fall trend is similar for all the pairs.
	
	\FloatBarrier

	%		\begin{figure}[h]
		%			\centering
		%			\includegraphics[width=0.7\textwidth]{apt vqsl.png}
		%			\caption{APT VQSL}
		%			\label{fig:center_trend_age}
		%		\end{figure}
	
	\section{Entanglement Entropy of the Rényi Group}
	
	We begin by reviewing the Rényi group of entropies, a one-parameter family of generalized entanglement measures \cite{renyi1961measures, nielsen2010quantum}. These entropies are especially useful in characterizing different spectral properties of the reduced density matrix \(\rho_s\).
	
	\subsection{Definition of the Rényi Entropy}\label{sec:bruh}
	
	For a normalized density matrix \(\rho_s\), the Rényi entropy of order \(q \in (0, \infty)-\{1\}\) is defined as:
	\begin{equation}
		S_q(\rho) = \frac{1}{1 - q} \log \mathrm{Tr}[\rho^q].
	\end{equation}
	which is a monotonically decreasing function of q.
	This function reduces to the Von Neumann entropy in the limit \(q \to 1\) \cite{nielsen2010quantum, muller2013quantum}:
	\begin{equation}
		S_1(\rho) = - \mathrm{Tr}[\rho \log \rho].
	\end{equation}
	Each value of \(q\) controls how sensitively the entropy measure reacts to large versus small eigenvalues of \(\rho\) \cite{muller2013quantum}. For \(q\)-values between 0 and 1, it emphasizes on smaller eigenvalues of $\rho$, while \(q\)-values higher than 1 emphasize the larger eigenvalues.
	
	We now explore the entropy at four distinguished values of \(q\), ordered from \(0\) to \(\infty\).
	
	\subsection{Rényi Entropy at Distinguished Orders}
	
	\subsubsection{\boldmath\(q = 0\): Max-Entropy}
	
	The Rényi entropy at \(q = 0\) captures the logarithmic rank of the density matrix:
	\begin{equation}
		S_0(\rho) = \log \mathrm{rank}(\rho)\label{eq:zeroentropy}.
	\end{equation}
	This quantity is insensitive to the actual eigenvalues and simply counts the number of non-zero ones, weighting all of their corresponding eigenstates with equal probabilities. It reflects the maximum possible entropy or largest possible uncertainty supported by the system and is sometimes called the "max-entropy" \cite{muller2013quantum}.

	\subsubsection{\boldmath\(q = 1\): Von Neumann Entropy}\label{sec:q1}
	
	The standard quantum entropy:
	\begin{equation}
		S_1(\rho) = - \mathrm{Tr}[\rho \log \rho].
	\end{equation}
	This measures the average uncertainty or entanglement entropy and is additive for independent systems, weighting all of their corresponding eigenstates linearly. It has deep connections to thermodynamics, quantum channel capacities, and entanglement theory \cite{nielsen2010quantum, breuer2002theory}.
	
	\subsubsection{\boldmath\(q = 2\): Collision Entropy}
	
	At \(q = 2\), we find:
	\begin{equation}
		S_2(\rho) = - \log \mathrm{Tr}[\rho^2].
	\end{equation}
	This entropy is directly related to the purity and discrimination between quantum states, and is frequently used to detect decoherence, quantify entanglement, and study complexity in many-body systems \cite{muller2013quantum}.
	
	\subsubsection{\boldmath\(q \to \infty\): Min-Entropy}
	
	Taking the limit \(q \to \infty\), the Rényi entropy becomes:
	\begin{equation}
		S_\infty(\rho) = - \log \lambda_{\max},
	\end{equation}
	where \(\lambda_{\max}\) is the largest eigenvalue of \(\rho\). This form, called the min-entropy, has operational meaning in single-shot information theory and bounds the optimal probability of correctly guessing the outcome of a measurement \cite{tomamichel2015quantum}.
	It represents the minimum surprise or maximum predictability.
	
	In cryptography, min-entropy measures how much randomness remains even if an adversary knows everything except the most likely outcome.
	\subsection{Summary}
	
	Different orders of the Rényi entropy highlight different aspects of a quantum state’s spectral profile. To summarize:
	
	\begin{center}
		\begin{tabular}{c|l|l}
			\(q\) & Name & Interpretation \\
			\hline
			\(0\) & Max-Entropy & \(\log\) of number of non-zero eigenvalues \\
			\(1\) & Von Neumann & Average uncertainty; thermodynamic entropy \\
			\(2\) & Quadratic Entropy & Related to purity; used in quantum chaos, decoherence \\
			\(\infty\) & Min-Entropy & \(-\log\) of maximum eigenvalue
		\end{tabular}
	\end{center}
	
	These Rényi entropies form a hierarchy:
	\[
	S_0(\rho)  \ge S_1(\rho) \ge S_2(\rho) \ge S_\infty(\rho).
	\]

	\subsection*{Zero order Rényi Entropy for \pt and \APT Symmetric Systems}
\begin{figure}[htbp]
		\centering
		\includegraphics[width=0.70\textwidth]{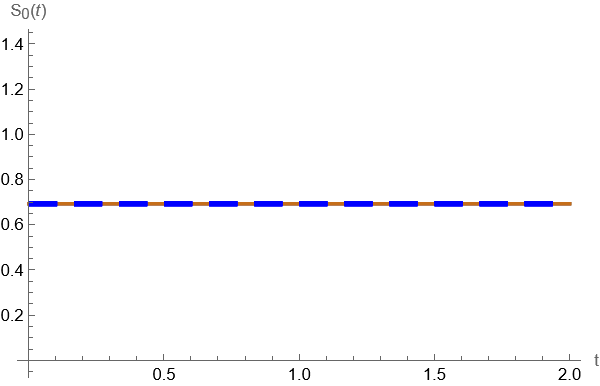}
		\caption{Zero-order Rényi Entropy: Blue represents \APT-symmetric systems and corresponding parameters are  $J_0 = 1$, $\beta = 0.5$, $\Omega_c = 1$, $\mu = -0.5$, $\theta = 0.86$, $\alpha = 1$. Orange shows \pt-symmetric systems and $J_0 = 1$, $\beta = 0.5$, $\Omega_c = 1$, $\mu = -0.5$, $\xi = 0.81$, $\alpha = 1$, $\delta = 0.56$. }
		\label{fig:entropy0}
	\end{figure}

	For both the \pt and Anti-\pt-symmetric systems, we can see from Fig.\ref{fig:entropy0} that this maximum entropy remains at the constant value 0.693 in time and independent of the non-Hermitian terms $\theta$ (for \pt) and $\xi$, $\delta$ (for \APT). Using Eq.\ref{eq:zeroentropy} for $S_{0}$, we can easily see that log(2) = 0.693 and thus the rank of the density matrix, and the dimension of the Hilbert space remain 2, implying that both systems are confined to stay in the two state space, and they cannot expand into any other subspaces during decoherence. However, we cannot attain any information on either the systems become a mixture of those two states, or how they weight and evolve between the two states or how those two states change in time.
	
	\FloatBarrier
	
	\vspace{3mm}
	\noindent
	\begin{flushleft}
		To demonstrate the decay of purity and entanglement with the environment, we further explore the higher orders of entropy in the following subsections.	
	\end{flushleft}
	\subsection*{First-order Rényi and the Von Neumann Entropy in \pt-Symmetric and \APT-Symmetric Qubits}
	
	To quantify the entanglement between the qubit and its surrounding environment, we employ the \textit{Von Neumann entropy}, a standard measure of quantum information and mixedness. Assuming an initially pure reduced density matrix with equal populations in the ground and excited states, the system begins in a maximally coherent superposition. 
	
	Under non-Hermitian dynamics, the reduced density matrix eigenvalues $\lambda_{1,2}(t)$ evolve over time, reflecting environmental influence. The Von Neumann entropy is then calculated as \cite{cen2022}
	\[
	S(t) = -\lambda_1(t)\ln\lambda_1(t) - \lambda_2(t)\ln\lambda_2(t).
	\]
	
	In our case, these eigenvalues depend on the decoherence function $D(t)$, resulting in the closed-form expression:
	\[
	S(t) = \ln 2 - \frac{1}{2}[1 + D(t)]\ln[1 + D(t)] - \frac{1}{2}[1 - D(t)]\ln[1 - D(t)].
	\]
	
	This expression reveals how entropy grows as coherence is lost, serving as a quantitative signature of entanglement and the irreversible nature of system-environment interactions.

\begin{figure}[htbp]
		\centering
		\hfill
		\begin{subcaptionbox}{\pt-Stmmetric System. $J_0 = 1$, $\beta = 0.5$, $\Omega_c = 1$, $\mu = -0.5$, $\xi = 0.81$, $\alpha = 1$, $\delta = 0.56$.\label{fig:firstpt}}[0.40\textwidth]
			{\includegraphics[width=\linewidth]{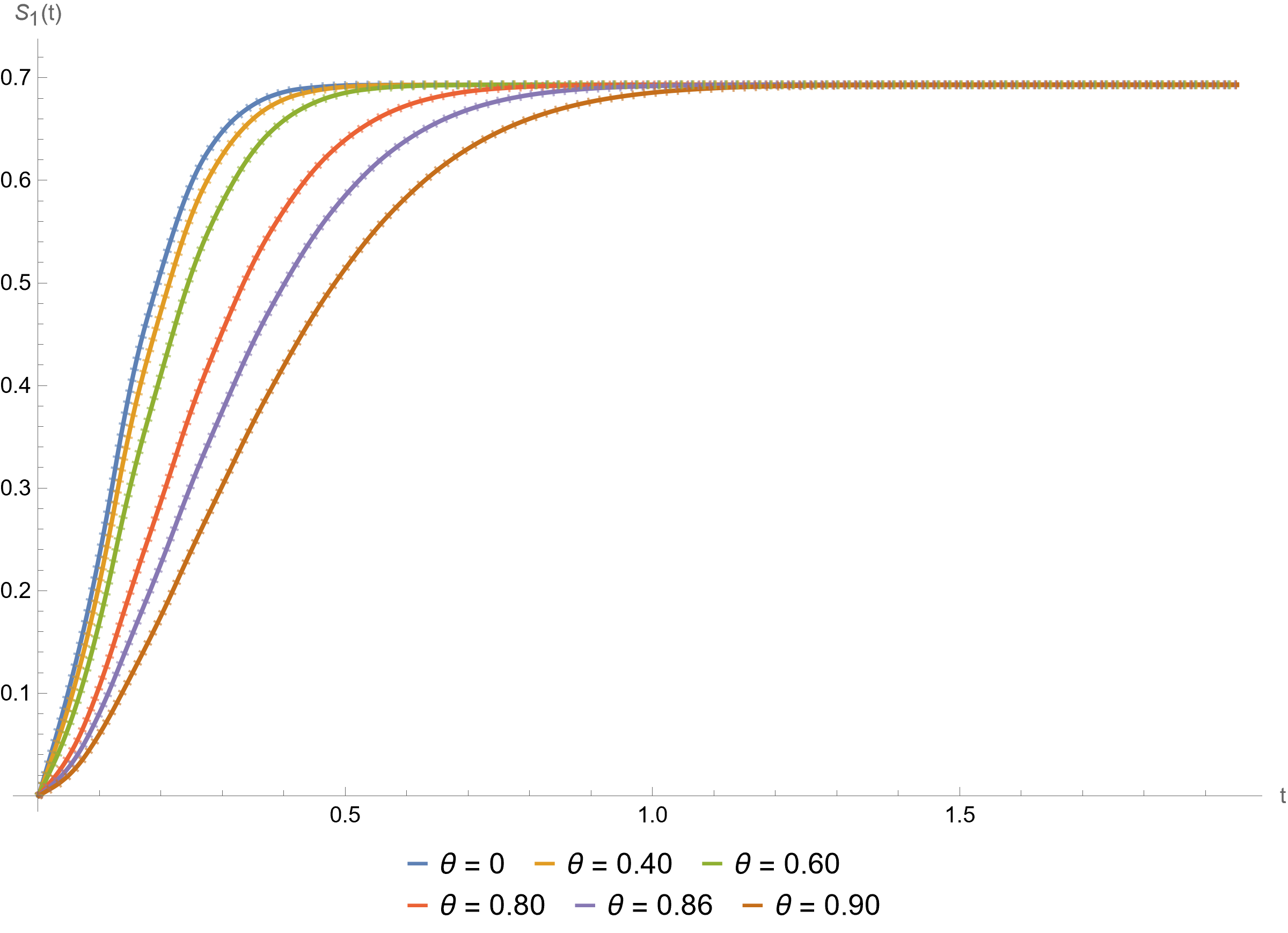}}
		\end{subcaptionbox}
		\hfill
		\begin{subcaptionbox}{\APT-Symmetric System. $J_0 = 1$, $\beta = 0.5$, $\Omega_c = 1$, $\mu = -0.5$, $\theta = 0.86$, $\alpha = 1$.\label{fig:firstapt}}[0.55\textwidth]
			{\includegraphics[width=\linewidth]{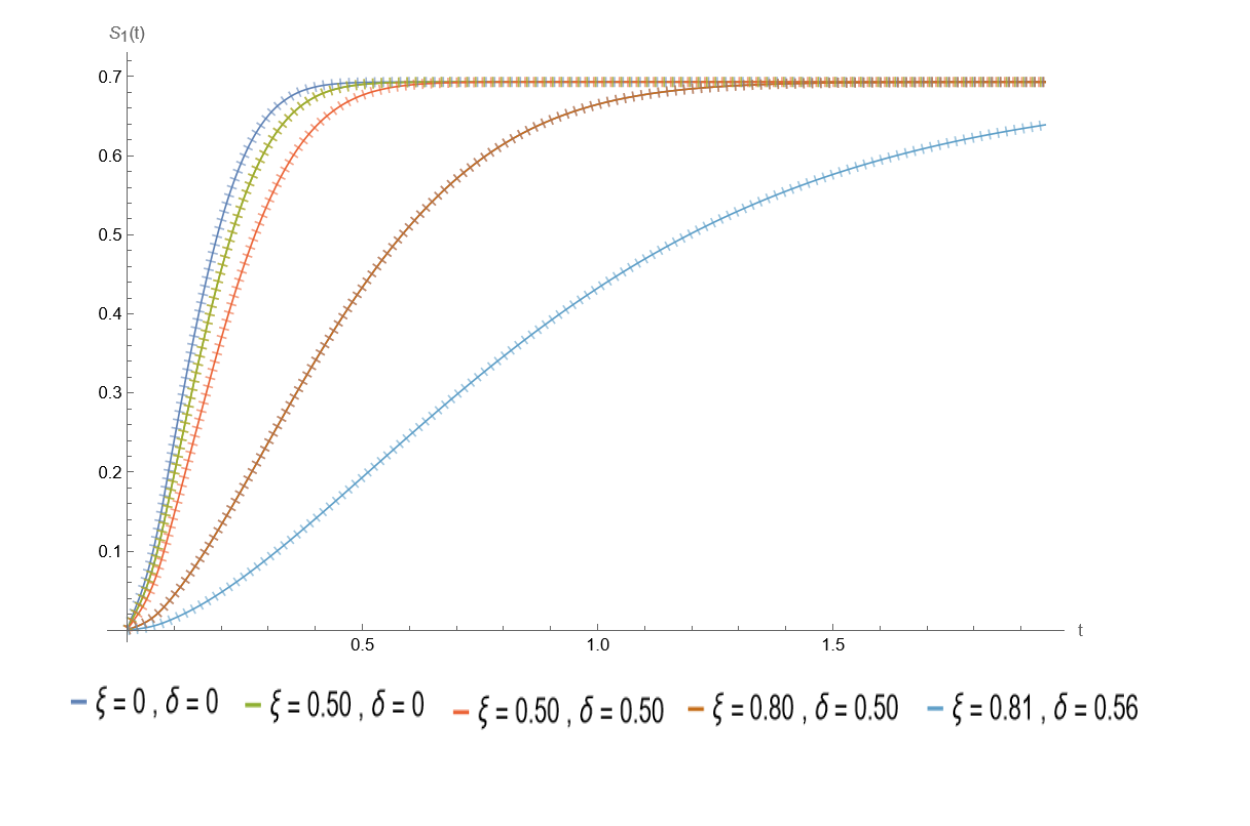}}
		\end{subcaptionbox}
		\caption{Solid curves represent the first order Rényi entropy and dotted curves exhibit Von Neumann entropy.
		}
		\label{fig:entropy1}
	\end{figure}

	As shown in section (\ref{sec:q1}), the 1st Reniy entropy reaches the Von Neumann entropy upon taking the limit \(q \to 1\). This can be verified for the \pt-symmetric and \APT-symmetric systems, as we have shown in figure \ref{fig:entropy1} .The solid curves, which are the the Reniy entropies of order 1, perfectly fit the dotted curves, which are the Von Neumann entropies. We can again imply that as non-Hermiticity increases for either of the systems, decoherence, leak of quantum information to the environment and entanglement with  the environment  slows down, as the plots become smoother for larger $\theta$ or $\xi$ and $\delta$ values. Also, this again perfectly aligns with the results shown in Fig. \ref{fig:pt_image2} and \ref{fig:apt_image2}. Notably, the best case for the \APT-symmetric system occurs in $\xi$ = 0.81, $\delta$ = 0.56 as suggested by \cite{cen2022}.
	
	\FloatBarrier
	
	\subsection*{Second-order (collision) Rényi Entropy in \pt-Symmetric and \APT-Symmetric Qubits}
\begin{figure}[htbp]
		\centering
		\hfill
		\begin{subcaptionbox}{\pt-Symmetric System. $J_0 = 1$, $\beta = 0.5$, $\Omega_c = 1$, $\mu = -0.5$, $\xi = 0.81$, $\alpha = 1$, $\delta = 0.56$.\label{fig:bruh}}[0.4\textwidth]
			{\includegraphics[width=\linewidth]{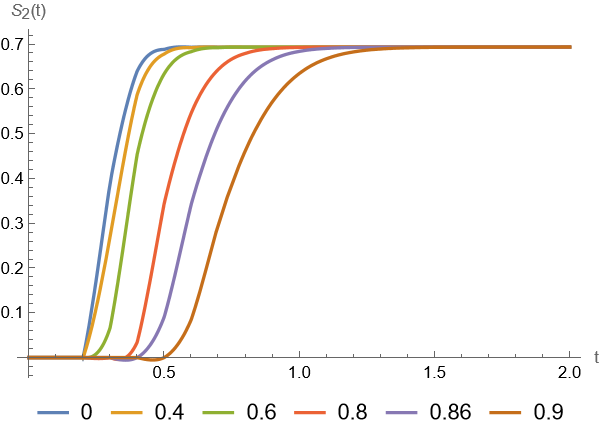}}
		\end{subcaptionbox}
		\hfill
		\begin{subcaptionbox}{\APT-Symmetric System. $J_0 = 1$, $\beta = 0.5$, $\Omega_c = 1$, $\mu = -0.5$, $\theta = 0.86$, $\alpha = 1$.\label{fig:img3}}[0.45\textwidth]
			{\includegraphics[width=\linewidth]{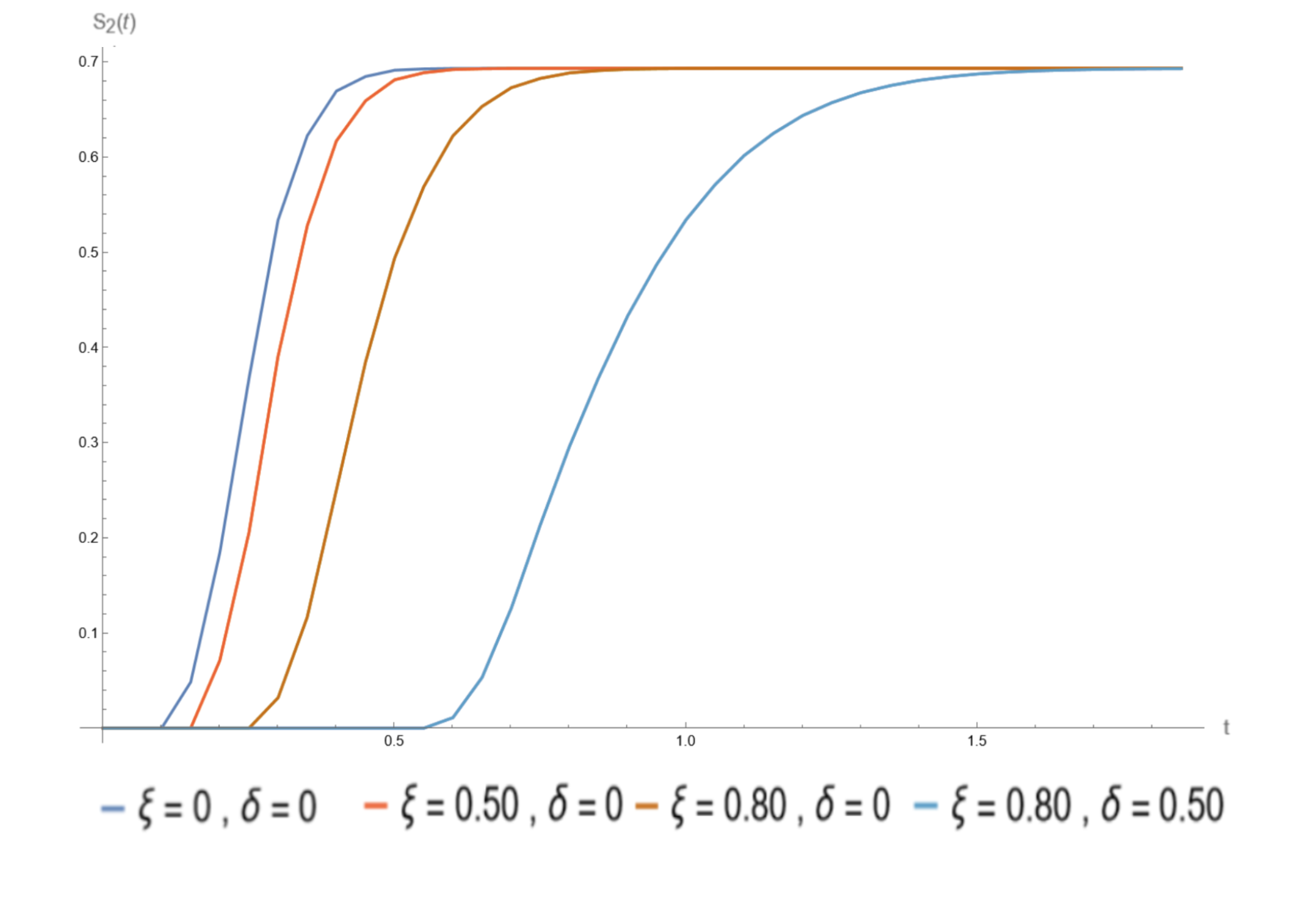}}
		\end{subcaptionbox}
		\caption{The 2nd Rényi entropy.}
		\label{fig:three}
	\end{figure}

	As shown in Fig. \ref{fig:three},  we observe that in both systems, the 2nd Rényi entropies are zero in the initial times but start to rise and finally saturate upon decoherence, with higher values of $\theta$ or ($\xi,\delta$) causing smoother behavior and slower rise.
	This is where collisions start. Each collision causes decoherence by transferring information from the system to the environment. The gradual rise reflects the cumulative effect of many small collisions.
	As the system evolves due to decoherence, the off-diagonal terms decohere, the state becomes mixed, and purity decreases.
	When the system reaches a steady mixed state, purity stabilizes, and entropy saturates to constant maximum for given Hilbert space.
	This corresponds to a steady mixed state where further collisions don’t increase entropy. The only difference between the systems is that in the \APT-symmetric case, the qubit with the pair ($\xi=0.8,\delta=0.5$) stays at zero entropy for longer and is more robust against decoherence, Fig. \ref{fig:img3}. 
	The ability of the \APT-symmetric qubit to remain in a low-entropy state for longer makes it a candidate for quantum memory applications, where maintaining quantum information with minimal loss is essential.

	\FloatBarrier
	
	\subsection*{Infinite-order (min) Rényi Entropy in \pt-Symmetric and \APT-Symmetric Qubits}

\begin{figure}[htbp]
		\centering
		\hfill
		\begin{subcaptionbox}{\pt-Symmetric System. $J_0 = 1$, $\beta = 0.5$, $\Omega_c = 1$, $\mu = -0.5$, $\xi = 0.81$, $\alpha = 1$, $\delta = 0.56$.\label{fig:fv2}}[0.4\textwidth]
			{\includegraphics[width=\linewidth]{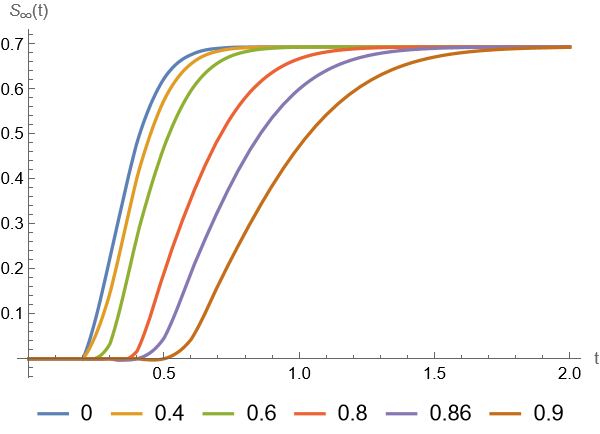}}
		\end{subcaptionbox}
		\hfill
		\begin{subcaptionbox}{\APT-Symmetric System. $J_0 = 1$, $\beta = 0.5$, $\Omega_c = 1$, $\mu = -0.5$, $\theta = 0.86$, $\alpha = 1$.\label{fig:fv1}}[0.45\textwidth]
			{\includegraphics[width=\linewidth]{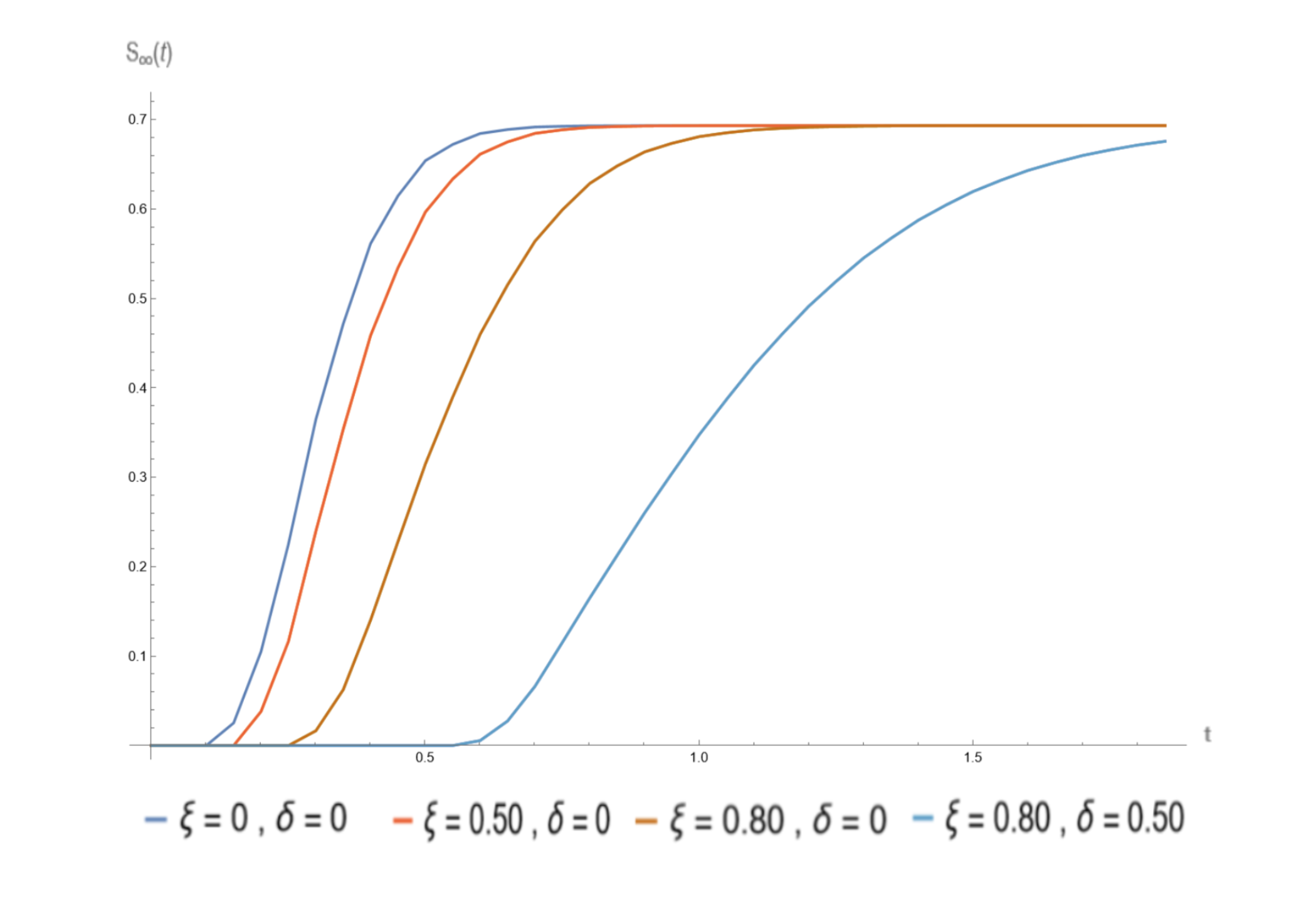}}
		\end{subcaptionbox}
		\caption{The $\infty$-order Rényi entropy}
		\label{fig:five}
	\end{figure}

	Similar to the second order, the observed behavior in Fig. \ref{fig:five} yields that initially, the min-entropy is zero, confirming one pure state where $\lambda_{\max}$ = 1, dominates completely. The rise in $S_\infty(\rho)$ reflects a decrease in $\lambda_{\max}$ as collisions redistribute probability among states, with smoother and slower rises for higher  $\theta$ or ($\xi, \delta$).  Saturation of $S_\infty(\rho)$ implies $\lambda_{\max}$ reaches a steady minimum, corresponding to a mixed state. For example, an n-dimensional Hilbert space corresponds to  $S_\infty(\rho) = n$, for a maximally mixed state ($\lambda_{\max} = \frac{1}{n}$).
	The min-entropy behavior, particularly the robustness of the pair ($\xi=0.8, \delta=0.5$), has significant applications in quantum cryptography, where prolonged low min-entropy enhances quantum key distribution security and ensures less information leakage to and correlation with the environment(asumming the environment as the eavesdropper).

	\FloatBarrier

	\section*{Conclusion}
	
	In this work, we conducted a thorough analysis of the dynamical behavior of qubits governed by \pt-symmetric and \APT-symmetric non-Hermitian Hamiltonians, focusing on the phenomena of phase evolution, decoherence, quantum speed limits (QSLs), and entropy evolution via the Rényi group.
	
	We observed that PT-symmetric systems, characterized by the commutation relation $[PT, H] = 0$, exhibit coherent-like dynamics within the unbroken symmetry phase, with real energy spectra and quasi-unitary evolution. Decoherence in these systems is modulated by the non-Hermitian parameter $\theta$, where an increase in $\theta$ generally decelerates the approach to mixed states. 
	
	\APT-symmetric systems, where $\{PT,H\}=0$, behave similar the \pt-symmetric systems. However, the key difference is that in \APT systems, $\xi$ and $\delta$ are the non-Hermitian parameters and govern the decoherence behavior \cite{cen2022}.
	
	As for the QSL, both systems show a transient increase before decreasing, suggesting initial acceleration in decoherence and gradual deceleration as systems evolve. In \pt-symmetric systems, higher $\theta$ values attribute to smoother behavior. In contrast, increase of the pair ($\xi$, $\delta$) in \APT-symmetric systems does not necessarily lead to smoother behavior.
	
	For both systems, Zero-order Rényi entropy remains constant in time regardless of the non-Hermitian terms and enforces both systems to stay in a two-dimensional Hilbert space. The first-order Rényi entropy reaches Von Neumann entropy in limit $q\to 1$, confirming previous analysis in section (\ref{sec:q1}). Both second and infinite order Rényi entropy behave similarly. In \APT-symmetric systems, a qubit with the pair ($\xi=0.8,\delta=0.5$) stays at zero entropy for the longest time in both orders, indicating slower decoherence.
	
	Overall, non-Hermitian systems with \pt and \APT symmetry tend to be more applicable in quantum information and related fields, for example quantum cryptography or computation, due to the their behavior. Systems with \APT symmetry outshine \pt-symmetric systems in performance under open-system dynamics and decoherence \cite{cen2022,junjie}, paving the way for more reliable resources for further applications.\\
	
	\section*{Acknowledgments}
We would like to thank Mina Ghodsi Yengejeh and Abasalt Rostami for thier valuable support and comments to this work.
			\bibliographystyle{apsrev4-2}
			\bibliography{references}
		\end{document}